\documentclass{article}
\usepackage{spconf,amsmath,graphicx,mathtools}
\usepackage{xcolor}
\usepackage{enumitem}
\setlist{nosep, leftmargin=14pt}

\usepackage{mwe} 
\usepackage{wrapfig}
\usepackage{url} 

\title{Three-dimensional diffusion-weighted multi-slab MRI with slice profile compensation using deep energy model}
\name{Reza Ghorbani*, Jyothi Rikhab Chand*, Chu-Yu Lee$\dagger$, Mathews Jacob*, Merry Mani*\thanks{This work is supported by NIH grants R01 EB031169, R01 EB031169-02S1, R01-AG067078, R01-EB019961.}}
\address{* University of Virginia, $\dagger$ University of Iowa }
\begin{document}
%
\maketitle
\begin{abstract}
Three-dimensional (3D) multi-slab acquisition is a technique frequently employed in high-resolution diffusion-weighted MRI in order to achieve the best signal-to-noise ratio (SNR) efficiency. However, this technique is limited by slab boundary artifacts that cause intensity fluctuations and aliasing between slabs which reduces the accuracy of anatomical imaging. Addressing this issue is crucial for advancing diffusion MRI quality and making high-resolution imaging more feasible for clinical and research applications. In this work, we propose a regularized slab profile encoding (PEN) method within a Plug-and-Play ADMM framework, incorporating multi-scale energy (MuSE) regularization to effectively improve the slab combined reconstruction. Experimental results demonstrate that the proposed method significantly improves image quality compared to non-regularized and TV-regularized PEN approaches. The regularized PEN framework provides a more robust and efficient solution for high-resolution 3D diffusion MRI, potentially enabling clearer, more reliable anatomical imaging across various applications.
\end{abstract}
\section{Introduction}

\label{sec:intro}

\noindent\ Three-dimensional multislab (3D multislab) acquisition has been employed in high resolution diffusion imaging as a signal-to-noise ratio (SNR) efficient alternative. In this approach, the entire 3D volume is segmented into multiple 3D sub-volumes along the slice direction, with each sub-volume referred to as a slab, where the slabs are independently excited and encoded \cite{engstrom2013diffusion}. However, the higher SNR efficiency offered by 3D multislab acquisition can be compromised by slab boundary artifacts, which result from imperfections in the radio frequency (RF) pulse profiles. The truncation of RF pulses leads to a slab-selective excitation profile that only approximates the ideal rectangular slab profile. This approximation introduces variations in the magnitude of the main lobe, along with nonzero transition bands and side lobes, contributing to the artifacts. The magnitude variation within the main lobe of the slab excitation profile results in signal intensity fluctuations in the reformatted image along the slab direction. Additionally, the presence of nonzero transition bands and side lobes causes the excitation profile to extend beyond the intended slab thickness, leading to two issues: first, slab crosstalk, where the imperfect slab profile excites portions of adjacent slabs, and second, aliasing artifacts.

Some earlier approaches mitigate slab boundary artifacts by oversampling in the slice direction, overlapping adjacent slabs, and combining the overlapped slices during reconstruction through methods such as averaging or cropping.  Slab profile encoding (PEN) is a recent approach that poses the slab boundary artifact correction similar to a standard MRI parallel imaging inverse problem without the need for oversampling. In this approach, corrected image is estimated from a linear system of of slab profile weighted images, where the slab profiles are assumed to be known from calibration scans and captures the side lobes and the uneven excitation profiles\cite{van2015slab}. 

Slab over-sampling and overlap leads to increased scan-time, thus reducing the SNR-efficiency of 3D multislab methods and restricting its utilization in high angular resolution applications. Correction of slab boundary artifacts from under-sampled 3D volumes helps to overcome this limitation. Here, we develop a regularized PEN reconstruction that can perform the slab boundary artifact correction from fewer samples. We formulate the regularized PEN as an iterative reconstruction that is solved in a Alternating Direction Method of Multipliers (ADMM) framework.

 Plug-and-Play ADMM is a widely adopted algorithm for addressing constrained optimization problems, particularly in image restoration tasks. One of its key advantages is its modular design, which enables the integration of any standard image denoising algorithm into a subproblem within the ADMM framework. Common image regularization employed in under-sampled recovery are Total Variation denoising or other pre-learned plug-and-play denoising. A problem with standard plug-and-play denoisers is the absence of convergence guarantees\cite{chan2016plug}. The recently proposed multi-scale energy (MuSE) framework ensures convergence guarantees while matching the performance of state of the art denoisers\cite{chand2024multi}. Here, we adopt the MuSE formulation as the regularizer in the plug and play ADMM framework.  We test the proposed formulation on 3D multislab data for slab combination and show improved performance compared to non-regularized and TV-denoised settings.

\section{Methods}

In this work, we present a Plug-and-Play ADMM approach to solve the regularized slab profile encoding problem. Our method leverages the properties of the MuSE regularization to correct the sharp transitions at slab boundaries caused by the non-regularized PEN approach. The following sections will discuss three key elements of our proposed method: the regularized objective function, the MuSE regularization, and the Plug-and-Play ADMM optimization framework.

\begin{figure}
    \centering 
    \includegraphics[width=1\columnwidth, height=0.7\columnwidth]{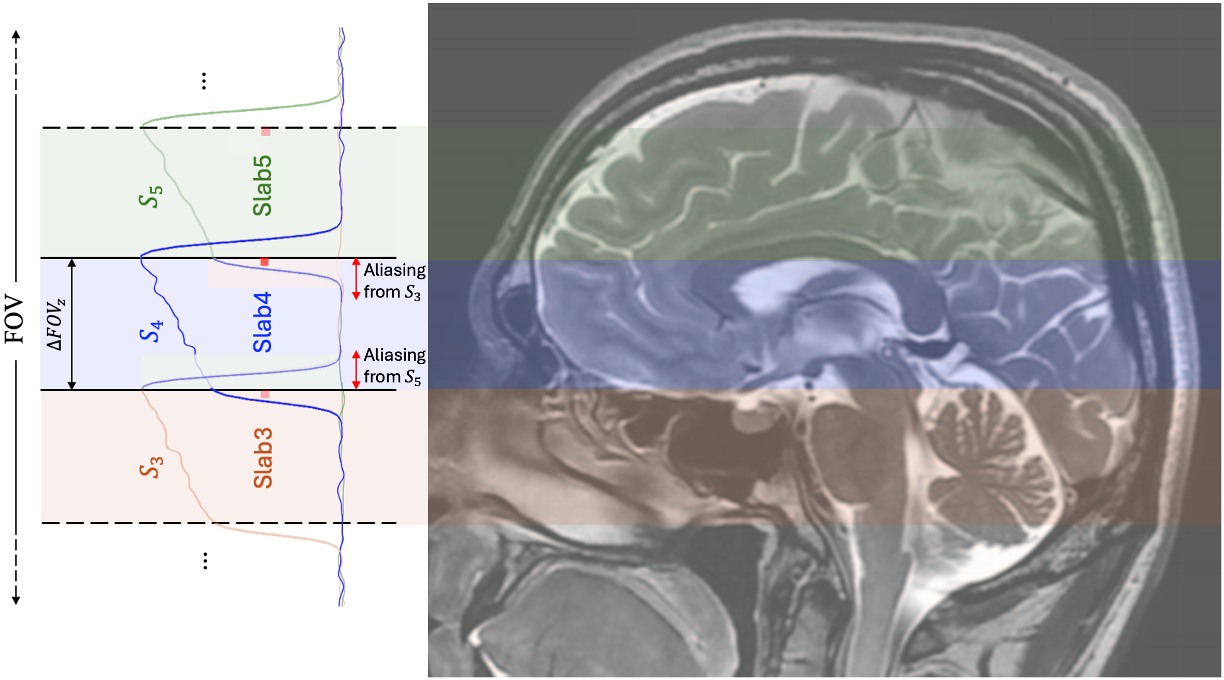}
    \vspace{-2.5em}
       \caption{Slab boundary artifact illustration.}
       \vspace{-1.5em}
    \label{fig:method}
\end{figure}

\label{sec:method}

\subsection{Problem Formulation}

Figure 1 illustrates the origin of the slab boundary artifact in a 3D multislab acquisition setup, where multiple slabs are acquired to cover the entire field of view (FOV). The ideal excitation thickness for each slab is denoted as $\Delta \text{FOV}_z$. If the excited area from the imperfect RF pulse exceeds $\Delta \text{FOV}_z$, aliasing will occur, where the signal at a given voxel within the intended thickness will receive contributions from locations outside the intended thickness $\Delta \text{FOV}_z$.

In the case of no over-sampling, the aliasing can be modeled mathematically as follows. Denoting $I_k(z)$ as the aliased signal measured at a voxel location $z$ for slab $k$, $I_k(z)$ is given by:
\vspace{-1.5em}
\begin{equation}
I_k(z) = ~~\smashoperator[lr]{\sum_{m=0}^{N_{slab} - 1}}~~S_{k}(z + m\Delta \text{FOV}_z)~ \rho(z + m\Delta \text{FOV}_z)
\end{equation}

\noindent for all $k$, $1\le$ $k\le$ $N_{slab}$, where $\text{FOV}_z$ is the full field of view along the slab direction, $N_{slab}$ is the total number of slabs required to cover the full $FOV_z, S_{k}$ is the pre-calibrated excitation profile of slab $k$, and $\rho$ is the unaliased 3D image volume that extends the full FOV, which needs to estimated.

The reconstruction problem in (1), in the presence of Gaussian noise $n$, can be compactly
represented in matrix notation as:
\begin{equation}
y = \mathcal{A}\rho + n 
\end{equation}
where $y$ is the measurement vector of the slab weighted images concatenated from all slabs $1\le k\le N_{slab}$, and $\mathcal{A}$ is the corresponding PEN operator that multiples the unaliased image volume with the slab profile weight at each voxel for a given slab $k$. In the maximum-a-posteriori (MAP) framework, the goal of the inverse problem is to maximize the posterior probability:

\begin{align}
\hat{\mathbf{\rho}} &= \arg \max_{\mathbf{\rho}} \, p(\mathbf{\rho} \mid \mathbf{y}) \notag \\
&= \arg \min_{\rho} \underbrace{ \| \mathbf{A}\rho - \mathbf{y} \|_2^2}_{\text{-log } p(\mathbf{y} \mid \rho)} - \log p(\rho) 
\end{align}

\noindent where the second equation also includes a log-prior term that serves as a regularization. While classical methods employ TV, low-rank priors etc as regularization, recent methods use data-driven deep-learned priors as regularization. In this
work, we employ a CNN-based prior represented as:
\begin{equation}
p_{\theta}(\rho) = \frac{1}{Z_{\theta}} \exp\left( {-\mathcal{E}_{\theta}(\rho)} \right)
\end{equation}

Here $\theta$ denotes the parameters of the CNN, $\mathcal{E}_{\theta}(\rho)$ denotes the energy function that is modeled by the CNN, and $Z_\theta$ is an unknown normalization constant. Given noise-free training data, the log-prior can be learned in a data-driven fashion using denoising score matching (DSM) as described in \cite{vincent2011connection}. The advantage of using DSM is that the gradient of the energy function is not dependent on the normalization constant, which is hard to evaluate.

\vspace{-1em}
\subsection{Energy Model in Multiple Dimensions}
\label{ssec:}
 Our goal is to minimize the residual artifacts in the slab boundaries while combining slab profile weighted images, which occur in the slice direction. Noise-free training data in this context is rare. Hence we train the CNN denoiser on axial slices without slab correction. The learned denoiser is then applied in the slices in the slab direction to remove noise while slab combination.  The energy is computed along three spatial directions x, y, and z of the MRI volume, as follows:
\begin{equation}
\mathcal{E}(\rho) = (\sum_x \mathcal{E} \left[ C_x(\rho) \right] + \sum_y \mathcal{E} \left[ C_y(\rho) \right] + \sum_z \mathcal{E} \left[ C_z(\rho) \right])/3
\end{equation}

Here $C_x$, $C_y$, $C_z$ are slice extraction operators along the $x$, $y$, $z$ directions, respectively. Because the extracted slices for a specific direction do not overlap, we have $\sum_x C_x^T C_x \rho = \rho$, which is also true for the other directions. The energy gradient or score $H_\theta(x)$ can be calculated using the chain rule or using PyTorch’s built-in autograd function. 

\vspace{-1em}
\subsection{MAP estimate using ADMM}

By substituting (5) in (4), equation (3) will be as follows:
\begin{equation}
\arg \min_{\rho} \| {A}\boldsymbol{\rho}- \mathbf{y} \|_2^2 + \lambda \mathcal{E}(\boldsymbol{\rho})
\end{equation}
 To solve (6) using ADMM, we start by defining
an auxiliary variables $\mathbf{v}$, and rewrite the cost function of (6) using this auxiliary variables as:
\begin{equation}
\arg \min_{\rho} \| {A}\boldsymbol{\rho}- \mathbf{y} \|_2^2 + \lambda \mathcal{E}(\mathbf{v}) \: s.t. \mathbf{v} =\boldsymbol{\rho}
\end{equation}
The augmented Lagrangian corresponding to the above constrained optimization problem is given by:
\begin{equation}
\mathcal{L}(\boldsymbol{\rho}, \mathbf{v}) = \| {A}\boldsymbol{\rho}- \mathbf{y} \|_2^2 + \lambda \mathcal{E}(\mathbf{v}) + \frac{\beta}{2} \|\mathbf{v}- \boldsymbol{\rho}\|^2+ \boldsymbol{\gamma}^T(\mathbf{v}- \boldsymbol{\rho})
\end{equation}
where $\boldsymbol{\gamma}$ is the Lagrange multiplier and $\beta>0$. To solve (8), each variable is sequentially updated by solving a set of sub-problems, during which the other variables are kept fixed:
\begin{equation}
\boldsymbol{\rho}^{n+1} = \arg \min_{\boldsymbol{\rho}} \| {A}(\boldsymbol{\rho})- \mathbf{y} \|_2^2 + \frac{\beta}{2} \Big\|\boldsymbol{\rho} - \left(\dfrac{\boldsymbol{\gamma}}{\beta} +\mathbf{v}^n \right)\Big\|^2 \tag{9.1}
\end{equation}
\begin{equation}
\mathbf{v}^{n+1} = \arg \min_{\mathbf{v}} \lambda  \mathcal{E}(\mathbf{v}) + \frac{\beta}{2} \Big\|\mathbf{v} - \left(\boldsymbol{\rho}^{n+1} - \frac{\boldsymbol{\gamma}}{\beta}\right)\Big\|^2 \tag{9.2}
\end{equation}
\begin{equation}
\boldsymbol{\gamma}^{n+1} = \boldsymbol{\gamma}^{n} + \beta \left(\mathbf{v}^{n+1} - \boldsymbol{\rho}^{n+1}\right) \tag{9.3}
\end{equation}
We solve (9.2) using steepest descent.

\section{Implementation details}

\subsection{Dataset}
 {In-vivo 3D-multislab diffusion MRI data was acquired with 8 slabs to cover the whole brain. Each slab covers 20 mm with 7 mm overlap. 10 diffusion weighted volumes and one non-diffusion weighted volume were collected with the following acquisition parameters: FOV(x-y): 210 mm, matrix size: 210×210, echo time (TE) = 70ms, repetition time (TR)=2s, using a 44-channel head coil. Other parameters include a partial Fourier of 0.7 and in-plane acceleration of 3. A 2D navigator echo was acquired during each TR, which was used in a phase-compensated iterative SENSE reconstruction\cite{pruessmann2001advances}. The images generated from the SENSE reconstruction of the 8 slabs  were used for testing the proposed method.

\subsection{Architecture and Training of Neural Network}
Noting that for the energy and score calculation, complex images represent as two-channel images (real and imaginary channels), we model the energy $\mathcal{E}_{\theta}: \mathbf{C}^m \rightarrow \mathbf{R}^+$ used in this paper as:
\begin{equation}
\mathcal{E}_{\theta}(\rho) = \frac{1}{2} \| \rho - \psi_{\theta}(\rho) \|^2,
\end{equation}
where, as we mentioned earlier, $\theta$ denotes the parameters of the energy function, and $\psi_{\theta}(\cdot): \mathbf{C}^m \rightarrow \mathbf{C}^m$ is a CNN network. If $\psi_{\theta}$ can be viewed as a denoiser, therefore, the energy can be interpreted as the magnitude of noise in the image $\rho$.

We represent $\psi_{\theta}(\rho)$ using a deep DRUNet \cite{zhang2021plug}. The four downscaling and upscaling layers in DRUNet, which include 64, 128, 256, and 512 channels, resemble a U-Net in construction. However, DRUNet differs from U-Net in that each downscaling and upscaling layer has four extra residual blocks. Between the 2×2 transposed convolution upscaling operation and the 2×2 strided convolution downscaling operation, each layer has a skip connection.  Then, we pre-trained the parameters of $\mathcal{E}_{\theta}(\rho)$ by selecting the standard deviations from a uniform distribution with a range of 0 to 0.1.
 \vspace{-1.05em}
\section{Experiments and results}

To test the performance of the proposed method, we compare it with the non-regularized PEN which is the baseline method without regularization. Also we compare our method with the TV-regularized PEN which incorporates the well-known Total Variation denoising to address artifacts. The TV-regularization was also implemented using the plug-and-play ADMM and regularization was applied in the axial, coronal and sagittal planes.

\begin{figure}[h!b]
    \centering 
    \includegraphics[width=1\columnwidth, height=0.8\columnwidth]{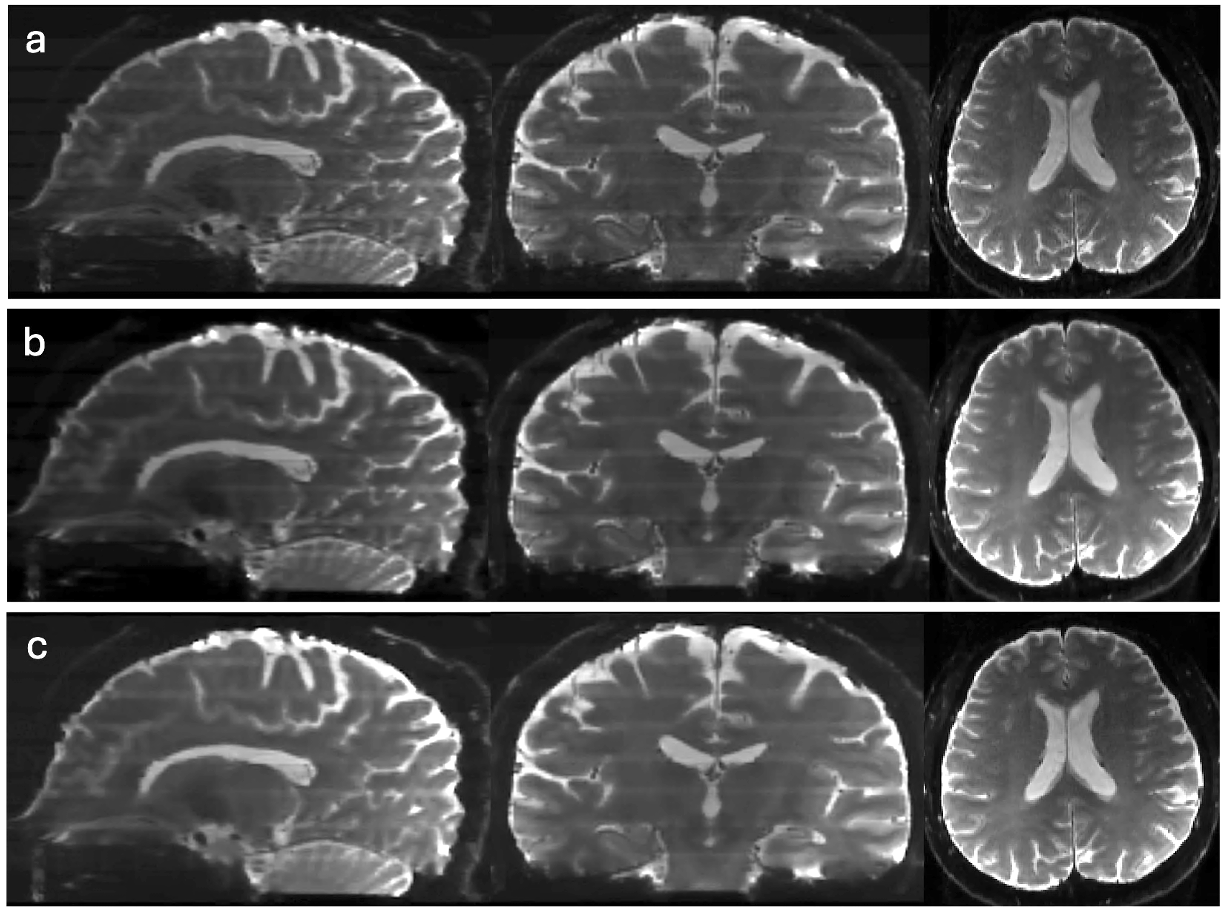}
    \vspace{-2.5em}
       \caption{T2 weighted images reconstructed using non-regularized PEN (a), TV-regularized PEN (b), and MuSE-regularized PEN (c).}
       \vspace{-0.7em}
    \label{fig:method}
\end{figure}

Figure 2 show the slab combined images of a non-diffusion weighted 3D volume using different slab combination methods. 
Figure 3 shows the reconstructed  diffusion-weighted volume using all the methods. The 10 diffusion volumes and the non-diffusion weighted volumes reconstructed using all the methods were used to test their reliability for diffusion tensor imaging (DTI). A tensor fitting was performed to estimate the fractional anisotropy (FA), primary diffusion direction, and mean diffusivity (MD) using the above data. The derived maps from the three methods are shown in Figure 4. 
It is noted that the regularized methods significantly improves the DTI fitting from fewer data compared to the non-regularized methods, with the MuSE-regularized PEN providing adequate denoising without oversmoothing compared to TV-regularized PEN. The boundary artifacts are visible in PEN and regularized PEN methods. This could be because of an imperfect RF pulse profile, whose artifacts are hard to correct while enforcing data consistency. Since the denoisers were not trained to correct such artifacts, this behaviour is expected. Additional datasets with improved RF pulses may help to suppress the boundary artifacts better.

\begin{figure}
    \centering 
    \includegraphics[width=1\columnwidth, height=0.77\columnwidth]{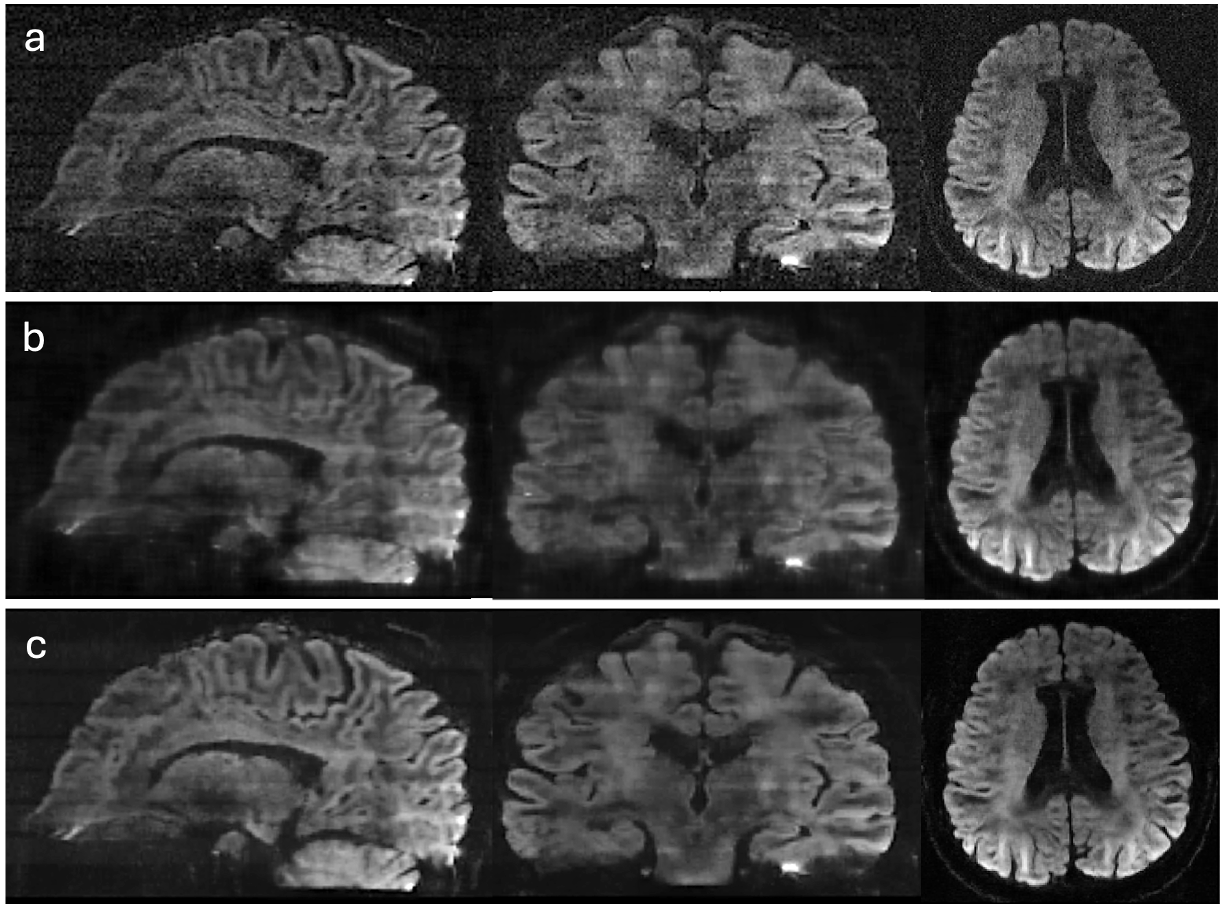}
     \vspace{-2.2em}
       \caption{Diffusion weighted images reconstructed using non-regularized PEN (a), TV-regularized PEN (b), and MuSE-regularized PEN (c).}
       \vspace{-0.8em}
    \label{fig:method}
\end{figure}

\begin{figure}
    \centering 
    \includegraphics[width=1\columnwidth, height=0.77\columnwidth]{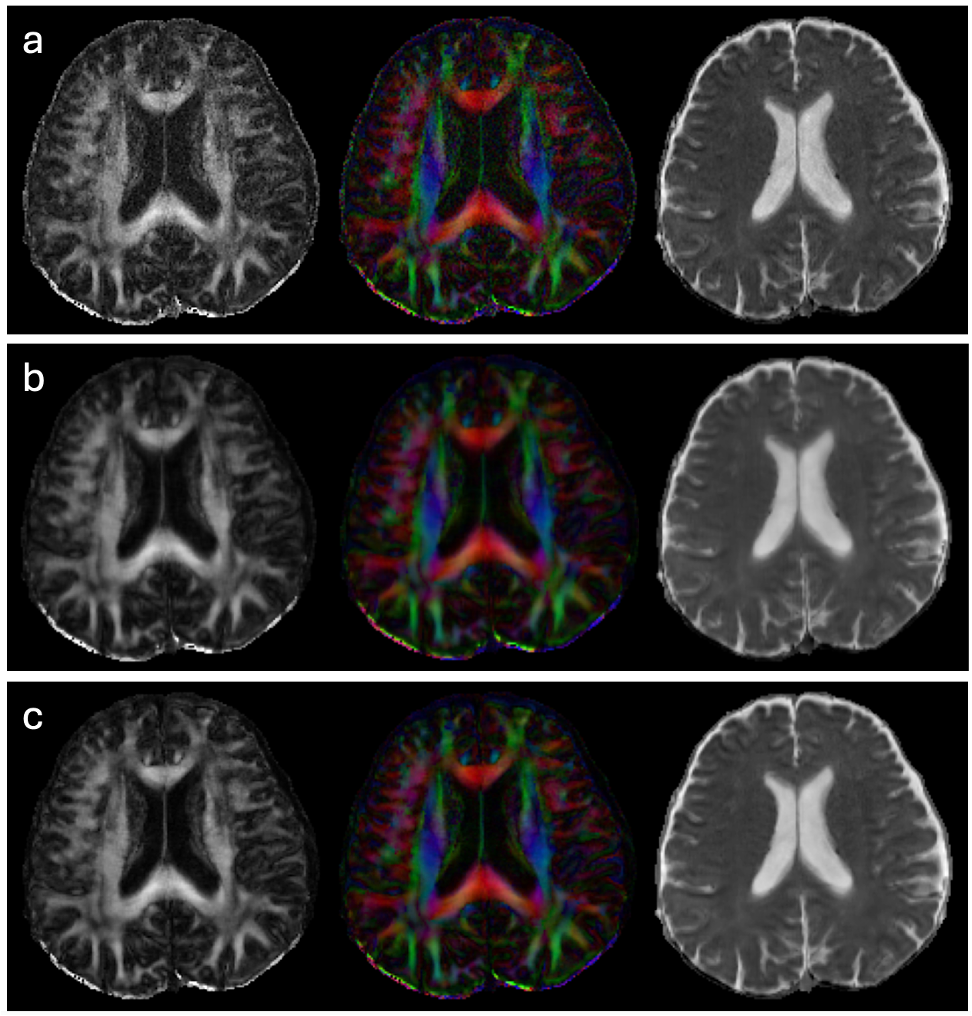}
     \vspace{-2.2em}
       \caption{FA (left), color-coded FA (middle), and MD (right)  maps reconstructed using PEN(a), TV-regularized PEN (b), and MuSE-regularized PEN (c).}
       \vspace{-1.2em}
    \label{fig:method}
\end{figure}

\vspace{-1em}
\section{Conclusion}
\vspace{-1em}
In this study, we developed a regularized slab profile encoding (PEN) framework for three-dimensional diffusion-weighted multi-slab MRI reconstruction, utilizing a Plug-and-Play ADMM approach with MuSE regularization. Our approach effectively denoises the reconstructed images making it a more robust option for high-resolution, SNR-efficient diffusion MRI applications. The residual slab boundary artifacts in the regularized and non-regularized reconstructions indicates an imperfect slab profile in the data consistency term, which can only be corrected by better design of RF pulses.  
\vspace{-1em}
\section{Compliance with ethical standards}
\vspace{-1em}
The data in this research study was acquired on a 3T GE Premier MRI scanner from a normal volunteer at the University of Iowa (UI). This study was approved by the institutional review board (IRB) at UI, and informed written consent was obtained from the volunteer prior to scanning. 


\bibliographystyle{IEEEbib}

\bibliography{strings}

\end{document}